\documentclass[twocolumn,showpacs,superscriptaddress,showkeys,reprint,amsmath,amssymb,aps,prd]{revtex4-1}

\usepackage{graphicx}
\usepackage{dcolumn}
\usepackage{bm}
\usepackage{tabularx}
\usepackage{amsmath}
\usepackage{enumitem}
\usepackage[bookmarks,bookmarksnumbered]{hyperref}
\hypersetup{colorlinks = true,
            linkcolor = blue,
            anchorcolor =red,
            citecolor = blue,
            pdfauthor=author}
\usepackage[capitalise]{cleveref}

\usepackage[dvipsnames]{xcolor}

\begin{document}

\title{Reconstructing spectral functions via automatic differentiation}

\author{Lingxiao Wang}
\affiliation{Frankfurt Institute for Advanced Studies, Ruth Moufang Strasse 1, D-60438,
Frankfurt am Main, Germany}

\author{Shuzhe Shi}
\email{shuzhe.shi@stonybrook.edu}
\affiliation{Department of Physics, McGill University, Montreal, Quebec H3A 2T8, Canada.}
\affiliation{Center for Nuclear Theory, Department of Physics and Astronomy, Stony Brook University, Stony Brook, New York, 11784, USA.}
\author{Kai Zhou}
\email{zhou@fias.uni-frankfurt.de}
\affiliation{Frankfurt Institute for Advanced Studies, Ruth Moufang Strasse 1, D-60438,
Frankfurt am Main, Germany}

\date{\today}

\begin{abstract}
Reconstructing spectral functions from Euclidean Green’s functions is an important inverse problem in many-body physics. However, the inversion is proved to be ill-posed in the realistic systems with noisy Green's functions. In this Letter, we propose an automatic differentiation(AD) framework as a generic tool for the spectral reconstruction from propagator observable. Exploiting the neural networks' regularization as a non-local smoothness regulator of the spectral function, we represent spectral functions by neural networks and use the propagator's reconstruction error to optimize the network parameters unsupervisedly. In the training process, except for the positive-definite form for the spectral function, there are no other explicit physical priors embedded into the neural networks. The reconstruction performance is assessed through relative entropy and mean square error for two different network representations. Compared to the maximum entropy method, the AD framework achieves better performance in the large-noise situation. It is noted that the freedom of introducing non-local regularization is an inherent advantage of the present framework and may lead to substantial improvements in solving inverse problems.
\end{abstract}

\maketitle

\emph{Introduction. }
The numerical solution to inverse problems is a vital area of research in many domains of science. In physics, especially quantum many-body systems, it’s necessary to perform an analytic continuation of function from finite observations which however is ill-posed~\cite{jarrell:1996bayesian,kabanikhin:2011inverse}. It is encountered for example, in Euclidean Quantum Field Theory (QFT) when one aims at rebuilding spectral functions based on some discrete data points along the Euclidean axis. More specifically, the inverse problem occurs when we take a non-perturbative Monte Carlo simulations (e.g., lattice QCD) and try to bridge the propagator data points with physical spectra~\cite{Asakawa:2000tr}. The knowledge of spectral function will be further applied in transport process and non-equilibrium phenomena in heavy ion collisions~\cite{Asakawa:2000tr,Rothkopf:2018jaj}. 

\begin{figure}[htbp!]
  \centering
  \includegraphics[width = 0.48\textwidth]{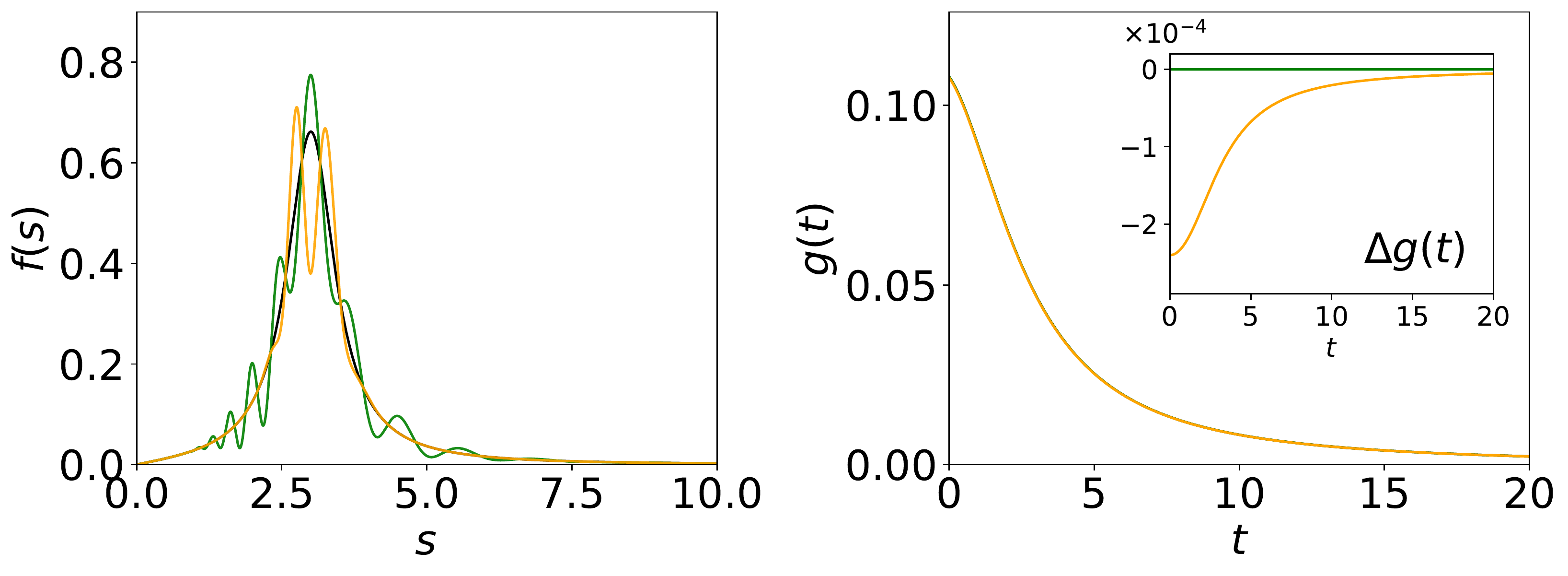}
  \caption{Spectral functions differed by null-modes (left) and their corresponding K\"allen--Lehmann correlation functions (right). The insert figure shows the differences-in-propagator caused by null-modes.}
  \label{fig:samples}
\end{figure}

In general, the problem set-up is from a Fredholm equation of the first kind, which takes the following form,
\begin{equation}
    g(t)=K \circ f:=\int_{a}^{b} K(t, s) f(s) d s,
\end{equation}
and the problem is to reconstruct the function $f(s)$ given the continuous kernel function $K(t,s)$ and the function $g(t)$. In realistic systems, $g(t)$ is often available in a discrete form numerically. When dealing with a finite set of data points with non-vanishing uncertainty, the inverse transform becomes ill-conditioned or degenerated~\cite{caudrey:1982inverse,tikhonov:1995numerical}. Regarding the convolution kernel as a linear operator, it can be expanded by basis functions in a Hilbert space. McWhirter and Pike~\cite{1978JPhA...11.1729M} and the authors~\cite{Shi:2022yqw} respectively show that kernels of Laplace transformation, ($K(t,s)=e^{-st}$), and K\"allen--Lehmann(KL) transformation, ($K(t,s)={s}({s^2 + t^2})^{-1}\pi^{-1})$,
have eigenvalues with arbitrarily small magnitude, and their corresponding eigenfunctions --- referred to as \textit{null-modes} --- induce negligible changes in function $g(t)$. Meanwhile, the {null-modes} correspond to arbitrarily large eigenvalues of the inversion operator. Therefore the inversion is numerically unstable when reconstructing $f(s)$ from a noisy $g(t)$. In Fig.~\ref{fig:samples}, we show examples of different $f(s)$ functions(at left hand side) that correspond to $g(t)$ functions with negligible differences(at right hand side).

\begin{figure*}[htbp!]
  \centering
  \includegraphics[width = 15 cm]{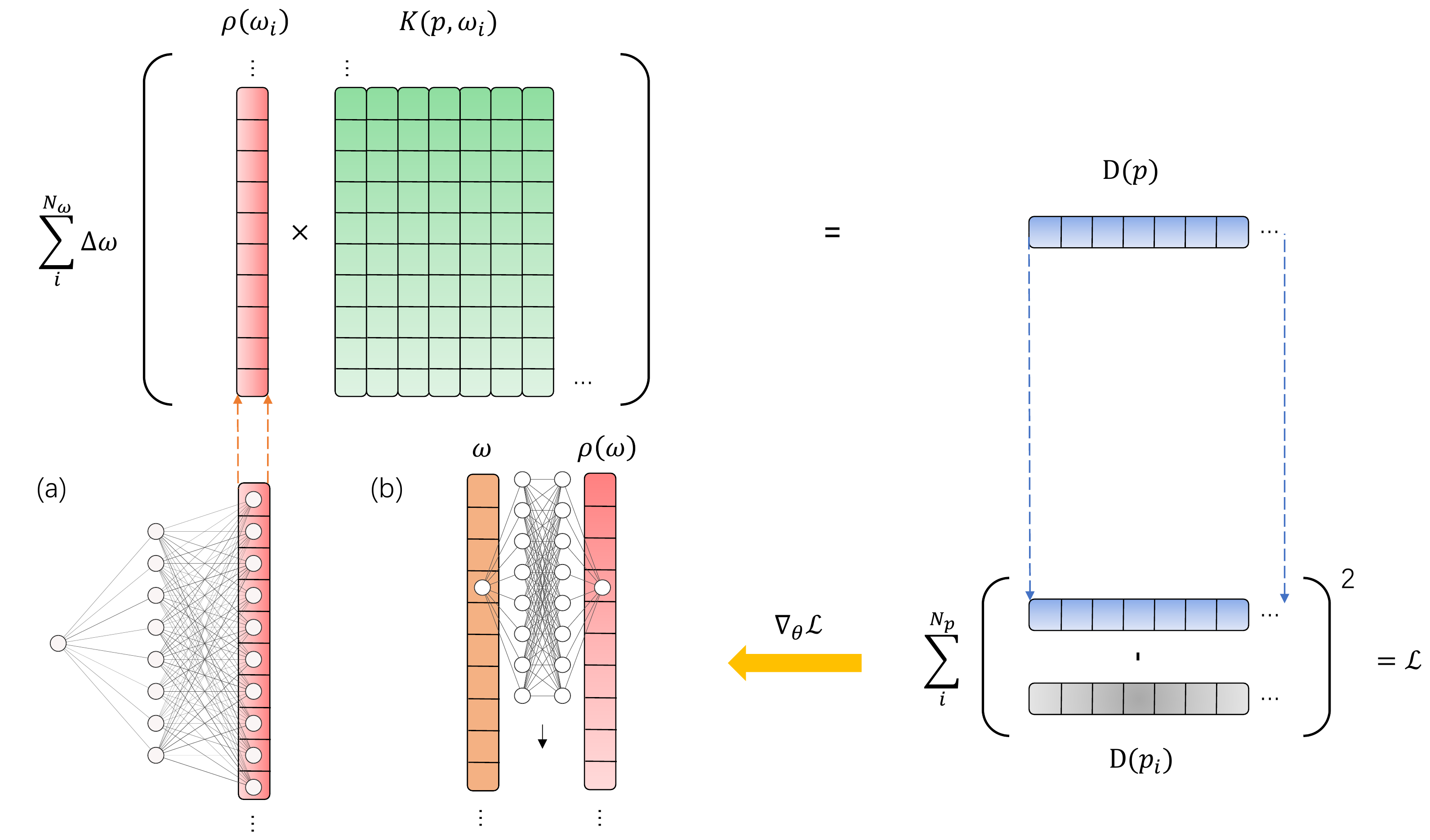}
  \caption{Automatic differential framework to reconstruct spectral from observations. (a) \texttt{NN}. Neural networks have outputs as a list representation of spectrum $\rho_i(\omega_i)$. (b) \texttt{NN-P2P}. Neural networks have input and output nodes as $(\omega_i,\rho_i)$ pairwise.
  \label{fig:frame}}
\end{figure*}

Many efforts have been made to break the degeneracy by adding regulator terms inside the inversion process, such as the Tikhonov regularization~\cite{bertero:1989linear,tikhonov:1995numerical}.
In past two decades, the most common approach in such reconstruction task is statistical inference. It comprises prior knowledge from physical domains to regularize the inversion~\cite{Asakawa:2000tr,Burnier:2013nla,Burnier:2014ssa}. As one classical paradigm, introducing Shannon--Jaynes entropy regularizes the reconstruction to an unique solution with suppressing null-modes~\cite{bryan:1990maximum,Asakawa:2020hjs,rothkopf2020bryan}, that is the maximum entropy method (MEM)~\cite{jarrell:1996bayesian,Asakawa:2000tr}. In general, the MEM addresses this problem by regularization of the least-squares fit with an entropy term $S[f]=-\int ds\left[f(s)-m(s)-f(s) \ln ({f(s)}/{m(s)})\right]$. Standard optimizations aim to maximize $Q[f]=\chi^{2}[f]/2-\alpha S[f]$ through changing $f(s)$ guided by a prior model $m(s)$, where $\alpha$ is a positive parameter that weights the relative importance between the entropy and the error terms. 
Although both Tikhonov and Shannon--Jaynes regularization terms yield unique solution of $f(s)$, it is not guaranteed that the reconstructed $f(s)$ is the physical one. Besides, there are some studies employing supervised approaches to train deep neural networks(DNNs) for learning the inverse mapping~\cite{Kades:2019wtd,2018PhRvB..98x5101Y,2020PhRvL.124e6401F,2020InvPr..36f5005L,Chen:2021giw}. In these works, the prior knowledge is encoded in amounts of training data from specific physics insights, whereas one should be careful about the risk that biases might be introduced in training sets. Efforts have been made in unbiased reconstructions by designing physics-informed networks and using complete basis to prepare training data sets~\cite{Chen:2021giw}. Besides, to alleviate the dependence on specific kinds of training data, there are also studies adopting the radial basis functions and Gaussian process~\cite{Zhou:2021bvw,Horak:2021syv} to perform the inversion directly.


In this Letter, we propose an unsupervised automatic differentiation(AD) approach to solve a spectral reconstruction task without training data preparation. Noting the oscillation caused by {null-modes}, it is natural to add smoothness condition to regularize the degeneracy. Therefore, we represent spectral functions by artificial neural networks(ANNs), in which the ANNs can preserve smoothness automatically\footnote{The universal approximation theorem ensures that ANNs can approximate any kind of continuous function with nonlinear activation functions~\cite{goodfellow:2016deep,2017arXiv170610239W}.}. Algorithms based on ANNs have been deployed to address various physics problems, e.g., determining the parton distribution function~\cite{Forte_2002, Collaboration_2007}, reconstructing the spectral function~\cite{Kades:2019wtd, Zhou:2021bvw,Chen:2021giw}, identifying phase transition~\cite{Carrasquilla_2017,Pang:2016vdc,Wang:2020hji, Wang:2020tgb, Jiang:2021gsw}, assisting lattice field theory calculation~\cite{PhysRevD.100.011501, Boyda:2020hsi, Kanwar:2020xzo, Albergo:2019eim}, evaluating centrality for heavy ion collisions~\cite{OmanaKuttan:2020brq, Thaprasop:2020mzp, Li:2020qqn}, parameter estimation under detector effects~\cite{Andreassen:2020gtw, Kuttan:2021npg}, and speeding up hydrodynamics simulation~\cite{Huang:2018fzn}.
Here we focus on the quality of the spectral function reconstructed from inverting the KL convolution~\cite{peskin:1995introduction},
\begin{align}
D(p) =\int_{0}^{\infty} K(p, \omega) \rho(\omega) d \omega \equiv \;& 
    \int_0^\infty 
    \frac{\omega \,\rho(\omega)}{\omega^2 + p^2} 
    \frac{d\omega}{\pi},
    \label{eq:corr}
\end{align}
where $D(p)$ is a propagator derived from a given spectral function $\rho(\omega)$. It is related to a wide range of quantum many-body systems, yet proved to be difficult to solve satisfactorily~\cite{rothkopf2020bryan,Asakawa:2020hjs}. It shall be worth noting that the framework discussed herein may be applied to other ill-conditioned kernels even extended to different tasks.

\begin{figure*}[htbp!]
    \centering
    \includegraphics[width = 0.98\textwidth]{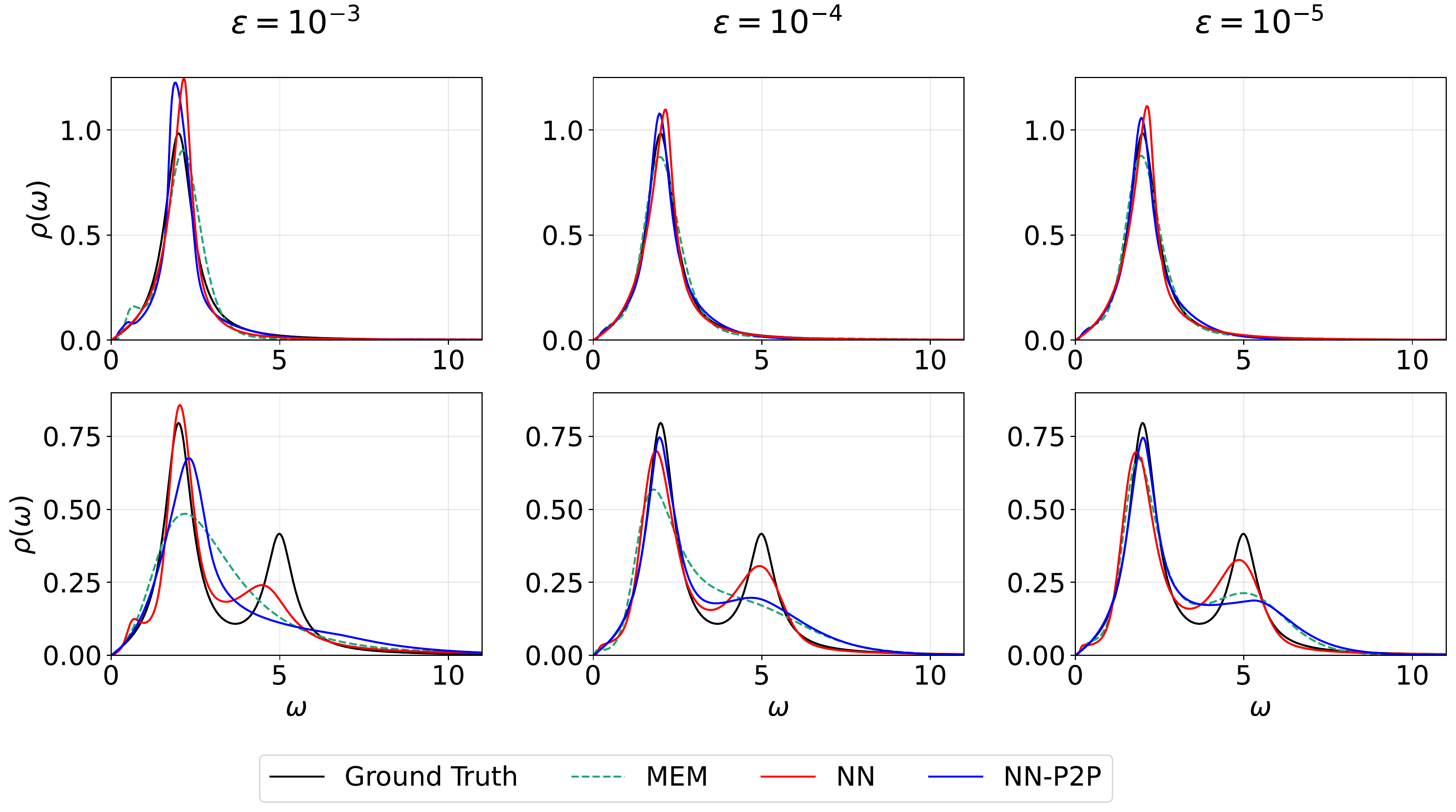}
    \caption{The predicted spectral functions from MEM, NN and NN-P2P. From left to right panels, different Gaussian noises are added to the propagator data with $\varepsilon = 10^{-3}, 10^{-4}$ and $10^{-5}$ in the case of $N_p = 25$, and $N_\omega = 500$ for the spectral. Note that MEM with fixed small $\alpha$ by hand might get improvement, as shown in Appendix with comparison to AD.
    \label{fig:rec}}
\end{figure*}

\emph{Automatic Differentiation. }
Fig.~\ref{fig:frame} shows the flow chart of the devised AD framework with network representations to reconstruct spectral from propagator observable. More details about the AD and related back-propagation algorithm can be found in Appendix. The output of network representations are $\vec{\rho}=[\rho_1,\rho_2,\cdots,\rho_{N_{\omega}}]$, from which we can calculate the propagator as $D(p)=\sum_i^{N_{\omega}} K(p,\omega_i) \rho_i \Delta \omega$. As Fig.~\ref{fig:frame} shows, after the forward process of the network and convolution, we can get the spectral $\vec{\rho}$ and further the correlators' reconstruction error as loss function,
\begin{equation}
    \mathcal{L}=\sum_i^{N_{p}} w_i (\mathrm{D}_i - D(p_i))^2,
    \label{eq.loss}
\end{equation}
where $\mathrm{D}_i$ is observed data at $p_i$, and $w_i$ denote extra weights of each observation. When taking the inverse variance as $w_i$, Eq.~\eqref{eq.loss} becomes the standard $\chi^2$ function. Meanwhile, one can directly extend it to multiple data points by making summation over them with calculating all variances. To optimize the parameters of network representations $\{\boldsymbol{\theta}\}$ with loss function, we implement gradient-based algorithms. It derives as,
\begin{align}
    \nabla_{\boldsymbol{\theta}} \mathcal{L} =
    \sum_{j,k}
    K(p_j,\omega_k)
    \frac{\partial \mathcal{L}}{\partial D(p_j)}
    \nabla_{\boldsymbol{\theta}} \rho_k,
\end{align}
where $\nabla_{\boldsymbol{\theta}} \rho_k$ is computed by the standard backward propagation(BP) method in deep learning~\cite{goodfellow:2016deep}. The reconstruction error will be transmitted to each layer of neural networks, combined with gradients derived from automatic differentiation~\footnote{It can be conveniently implemented in many deep learning frameworks. In our case, main computations are deployed in Pytorch and released on \href{https://github.com/Anguswlx/NNSpectrum.git}{Github}, but also validated in Tensorflow.}, they are used to optimize the parameters of neural networks. In our case, the Adam optimizer is adopted in following computations~\footnote{It is a stochastic gradient-based algorithm that is based on adaptive estimations of first-order and second-order moments~\cite{2014arXiv1412.6980K}.}. 

\emph{Neural network representations. }
As Fig.~\ref{fig:frame} shown, we develop two representations with different levels of non-local correlations among $\rho(\omega_i)$'s to represent the spectral functions with artificial neural networks(ANNs). The first is demonstrated as Fig.~\ref{fig:frame}(a) and named as \texttt{NN}, in which we use $L$-layers neural network to represent in list format the spectral function $\rho(\omega)$ with a constant input node $a^0=C$ and multiple output nodes $a^L = [\rho_1,\rho_2, \cdots,\rho_{N_{\omega}}]$. The width of the $l$-th layer inside the network is $n_l$, to which the associated weight parameters control the correlation among the discrete outputs in a concealed form. In a special case, a discrete list of $\rho_i$ itself is equivalent to set $L = 1$ without any bias nodes, meanwhile, the differentiable variables are directly elements of $\vec{\rho}$ as network weights. If one approximates the integration over frequencies $\omega_i$ to be summation over $N_\omega$ points at fixed frequency interval $d\omega$, then it is suitable to the vectorized AD framework described above. The second representation with ANNs is shown as Fig.~\ref{fig:frame}(b), where the input node is $a^0= \omega$ and the output node is interpreted to be $a^L = \rho(\omega)$. It is termed as point-to-point neural networks (abbreviated as \texttt{NN-P2P}) and it consists of finite first-order differentiable modules, in which the continuity of function $\rho(\omega)$ is naturally preserved~\cite{2017arXiv170610239W,rosca:2020case}.

We adopt $\text{width} = 64$ and $\text{depth} = 3$ as default parameter setting in the whole paper, which is explained in Appendix. Besides, a pedagogical introduction of the machine learning background can also be found there. For the optimization of the neural network representations, we adopt the Adam optimizer~\cite{kingma2014adam} with $L_2$ regularization for \texttt{NN}, which is a summation over the $L_2$ norm of all differentiable weights of the network, $L_2 = \lambda \sum_i (\theta_{W,i})^2$ with $\lambda = 10^{-2}$ in the beginning of the warm-up stage of training process. For speeding up the training process, we obey an annealing strategy to loosen the value of $\lambda$ from the initial tight regularization repeatedly to small enough value (smaller than $10^{-8}$) in first 40000 epochs. We checked that end values of $\lambda$ do not alter the reconstruction results once it's smaller than $10^{-8}$.
To converge fast, we also adopt a smoothness regulator here, which derives as $L_s = \lambda_s \sum_{i=1}^{N_{\omega}}(\rho_i- \rho_{i-1})^2$. The initial smoothness regulator is $\lambda_s = 10^{-3}$, then it decreases to 0 in the final step of the warm-up.  After that, early stopping is applied for the training with the criterion to be when error between observed $\tilde{D}(p)$ and reconstructed $D(p)$ does not decrease, or the whole training exceeds 250000 epochs. The learning rate is $10^{-3}$ for all cases, and there is no any explicit regulators for \texttt{NN-P2P} only implicit non-local correlations. Besides, the physical prior we embedded into these representations is the positive-definiteness of fermionic spectral functions (in Lattice QCD case, they are hadronic spectra), which is introduced by applying \textit{Softplus} activation function at output layer as $\sigma(x)=\ln(1+e^x)$.

\emph{Reconstruction performance. }
In this section, we demonstrate the performance of our framework by testing their quality in reverting the Green's functions of known spectral functions (aka. mock data). We start with a superposed collection of Breit--Wigner peaks, which is based on a parametrization obtained directly from one-loop perturbative quantum field theory~\cite{Tripolt:2018xeo,Kades:2019wtd}. Each individual Breit--Wigner spectral function is given by,
\begin{equation}
    \rho^{(\mathrm{BW})}(\omega)=\frac{4 A \Gamma \omega}{\left(M^{2}+\Gamma^{2}-\omega^{2}\right)^{2}+4 \Gamma^{2} \omega^{2}}.
    \label{eq:bw}
\end{equation}
Here $M$ denotes the mass of the corresponding state, $\Gamma$ is its width and $A$ amounts to a positive normalization constant. The multi-peak structure is built by combining different single peak modules together. 

Two profiles of spectral functions from Eq.~\eqref{eq:bw} are set as ground truths. In Fig.~\ref{fig:rec}, the upper is from a single peak spectrum with $A= 1.0, \Gamma = 0.5, M= 2.0$ (hereunder in paper, we omit the energy unit of mass $M$, width $\Gamma$, frequency $\omega$ and momentum $p$) and the below one is from double peak profile with $A_1= 0.8, A_2 = 1.0, \Gamma_1 = \Gamma_2 = 0.5, M_1= 2.0, M_2 =5.0$.  To imitate the realistic observable data, we follow Ref.~\cite{Asakawa:2000tr} and add noise to the mock data with $\Tilde{D}_i = \mathrm{D}(p_i) + n_{i,\varepsilon}$, where the noise term follows normal distribution with variance $\sigma_{i,\varepsilon}^2 = (\varepsilon\, \mathrm{D}(p_i) p_i/\Delta p_i)^2$, $P(n_{i,\varepsilon}) = \mathcal{N}(0, \sigma_{i,\varepsilon}^2)$. In Fig.~\ref{fig:rec}, we compare the reconstruction results with $\varepsilon=10^{-3}$, $10^{-4}$, $10^{-5}$, respectively. The two network representations are marked by blue and red lines. They all show remarkable reconstruction performances for a single peak at each noise level. As a comparison, results from MEM
are also shown as green lines. We see that, MEM show oscillations around zero-point under different noise backgrounds. The rebuilding spectral function from \texttt{NN-P2P} do not oscillate. This is especially important for such a task of extracting the transport coefficients from real-world lattice calculation data~\cite{Asakawa:2000tr,Kades:2019wtd}. 

For mock data with two peaks, we observe that the non-local smoothness condition of \texttt{NN-P2P} slightly suppress the bimodal structure, whereas \texttt{NN} successfully unfolds the two peaks information from Green's functions even with noise $\varepsilon = 10^{-3}$. Although \texttt{NN-P2P} misses the second peak which may appear in the case of bimodal as MEM, the calculations of different order momentum from spectral function will not be disturbed. Another advantage of the \texttt{NN-P2P} architecture is its stable performance of the spectral function at small $\omega$ limit, which is important for the measurement of conductivity $\sigma \propto \lim_{\omega\to0}{\rho(\omega)}/{\omega}$~\cite{Ding:2015ona,Ratti:2018ksb}. The smoothness condition automatically encoded in the network set-up suppresses the oscillating null-modes especially at small frequency region, and therefore allows the reliable extraction of conductivity in \texttt{NN-P2P}.

\begin{figure}[htpb!]
    \centering
    \includegraphics[width =0.48\textwidth]{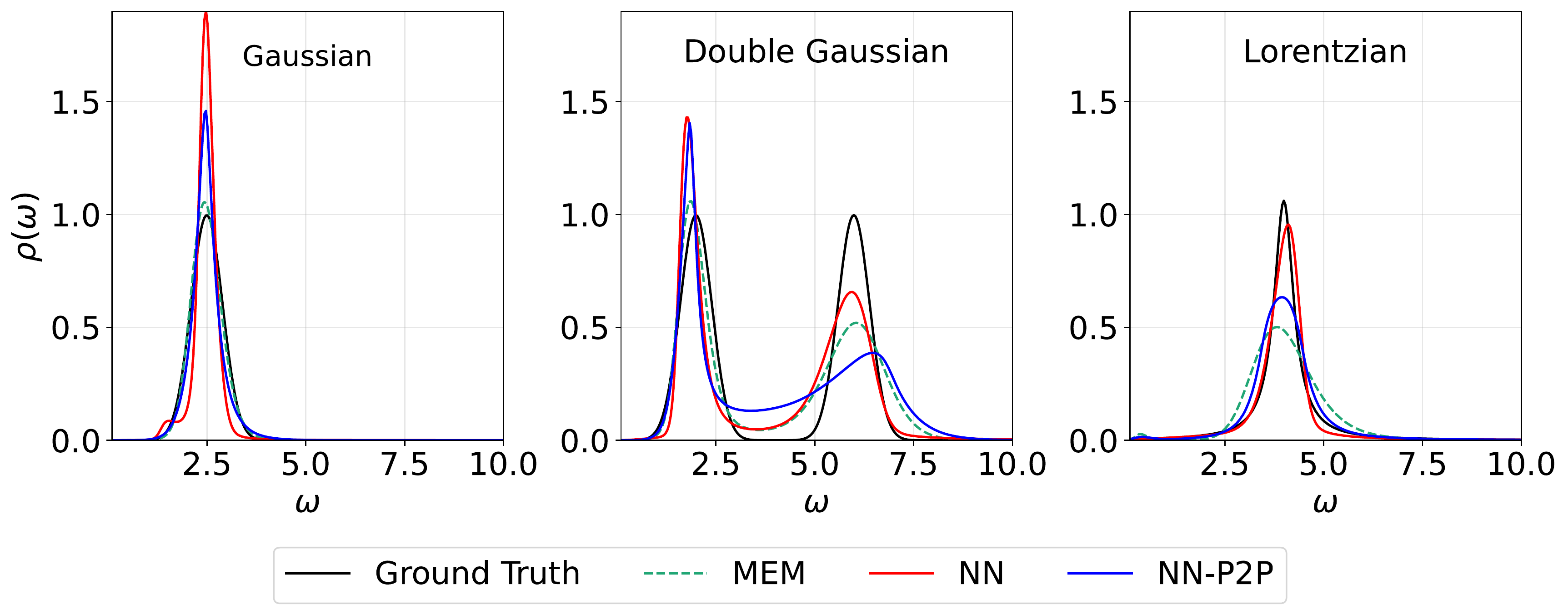}
    \caption{Reconstructed spectra with MEM, \texttt{NN} and \texttt{NN-P2P} from the correlators at noise level $\epsilon = 10^{-4}$ with $N_p = 25$ points.
    \label{fig:others}}
\end{figure}

In order to examine the robustness of our method against forms of spectral function, we further apply the framework to mock data prepared by \textit{Gaussian} form $\rho_G(\omega)= (2\pi\Gamma^2)^{-1}\exp{-((\omega - M)^2/2\Gamma^2)}$ (A single peak spectrum with $\Gamma = 0.4, M= 2.5$ and the double peak profile is setting as $\Gamma_1 = \Gamma_2 = 0.4, M_1= 2.0, M_2 =6.0$), and \textit{Lorentzian} form $\rho_L(\omega)= \Gamma^2[\pi\Gamma((\omega - M)^2 + \Gamma^2)]^{-1}$ with  $\Gamma = 0.3, M= 4.0$. The results are shown in Fig.\ref{fig:others}, and it indicates that neural network representations, \texttt{NN-P2P} and \texttt{NN}, can be generalized to other cases, and can reach at least comparable performances to the MEM method.

\emph{Extensions. } In addition to the above reconstructions, we also validate the framework in another two physics motivated cases. The first is to rebuild non-positive-definite spectral functions -- where classical MEM approaches are normally not applicable, unless adopting suitable representations within Bayesian Inference perspective~\cite{Hobson:1998bz,Burnier:2013nla,Rothkopf:2016luz,Horak:2021syv}. The reason for choosing such a set-up is that there are many circumstances the spectra would display positivity violation, which can be related to confined particles e.g., gluons and ghosts, or thermal excitations with long-range correlation in strongly coupled system~\cite{Rothkopf:2016luz,Dudal:2019gvn,Horak:2021syv}.
\begin{figure}[ht!]
    \centering
    \includegraphics[width =0.48\textwidth]{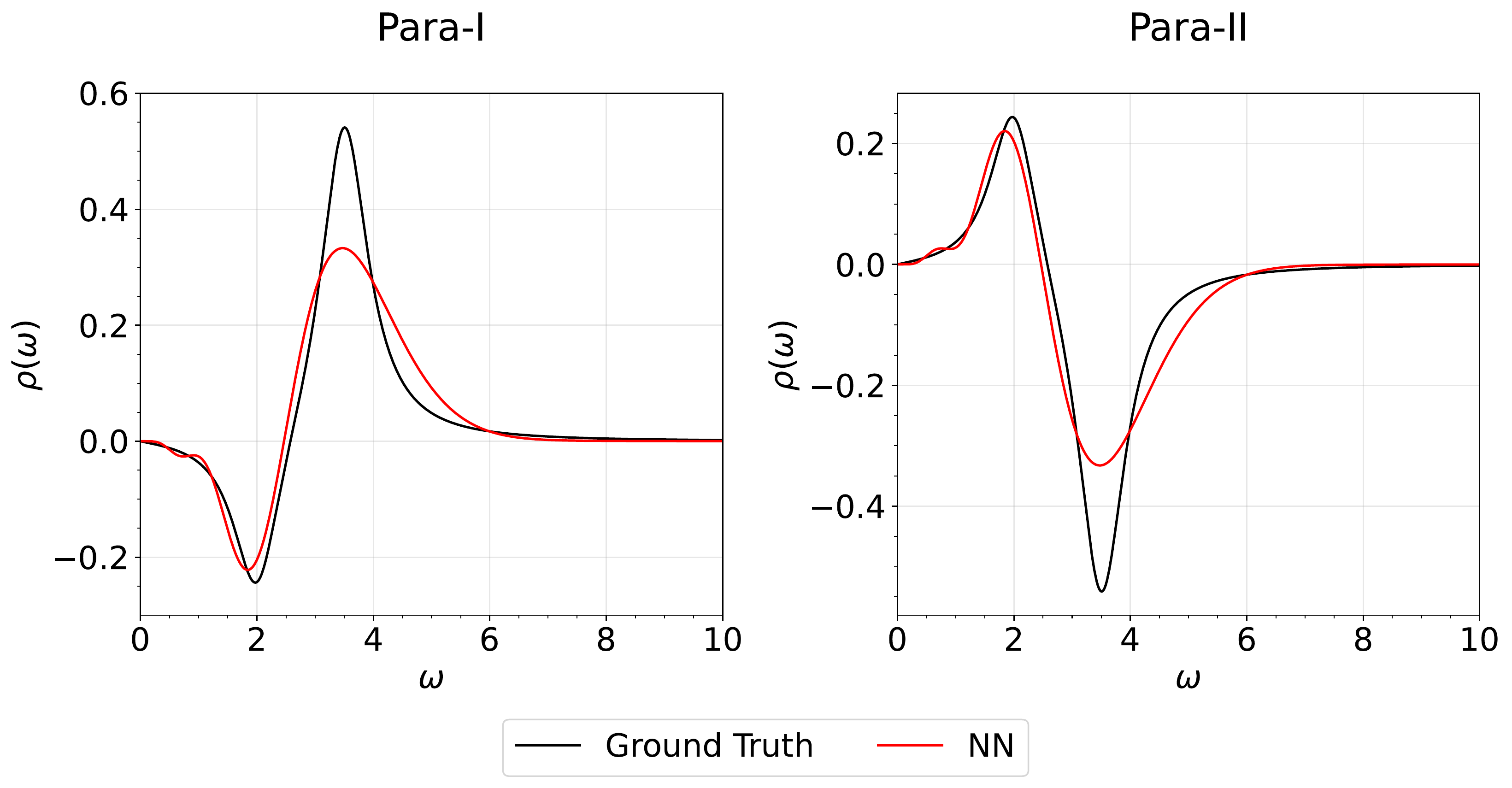}
    \caption{The predicted spectral functions from \texttt{NN}. The correlators are reconstructed at noise level $\epsilon =10^{-4}$ with $N_p = 25$ points.
    \label{fig:gluon}}
\end{figure}
In Fig.~\ref{fig:gluon}, we test our \texttt{NN} representation using the correlators generated from the double peak profile, with the first peak turning negative, $A_1= -0.3, A_2 = 1.0, \Gamma_1 = \Gamma_2 = 0.5, M_1= 2.0, M_2 =3.5$(Para-I) and $A_1= 0.3, A_2 = -1.0, \Gamma_1 = \Gamma_2 = 0.5, M_1= 2.0, M_2 =3.5$(Para-II). The errors added to correlators obey the same form explained before. The hierarchical architecture of \texttt{NN} representation is unchanged, but the positive activation function of output layer is removed to loosen the positive-definite condition, accordingly the multiplier factor is replaced by $\omega\,e^{-\omega}$ to suit the low- and large-$\omega$ limits. The reconstructions indicate that our NN works consistently well in constructing such spectral functions with non-positive parts at the location and width of peaks.

The other demonstration case we did is in a more realistic scenario. The hadron spectral function $\rho(\omega,T)$ is encoded in a thermal correlator $G(\tau,T)$ at temperature $T$~\cite{Tripolt:2018xeo}. The temperature dependent correlator can be calculated along the imaginary time $\tau$-axis. The physics motivated spectral functions proposed in Ref.~\cite{Chen:2021giw} are used to test our framework. The correlators are generated with Lattice QCD noise-level noises. The spectral function has two parts, a resonance peak and a continuum function. Details can be found in our Supplemental Materials.
\begin{figure}[ht!]
    \centering
    \includegraphics[width = 0.48\textwidth]{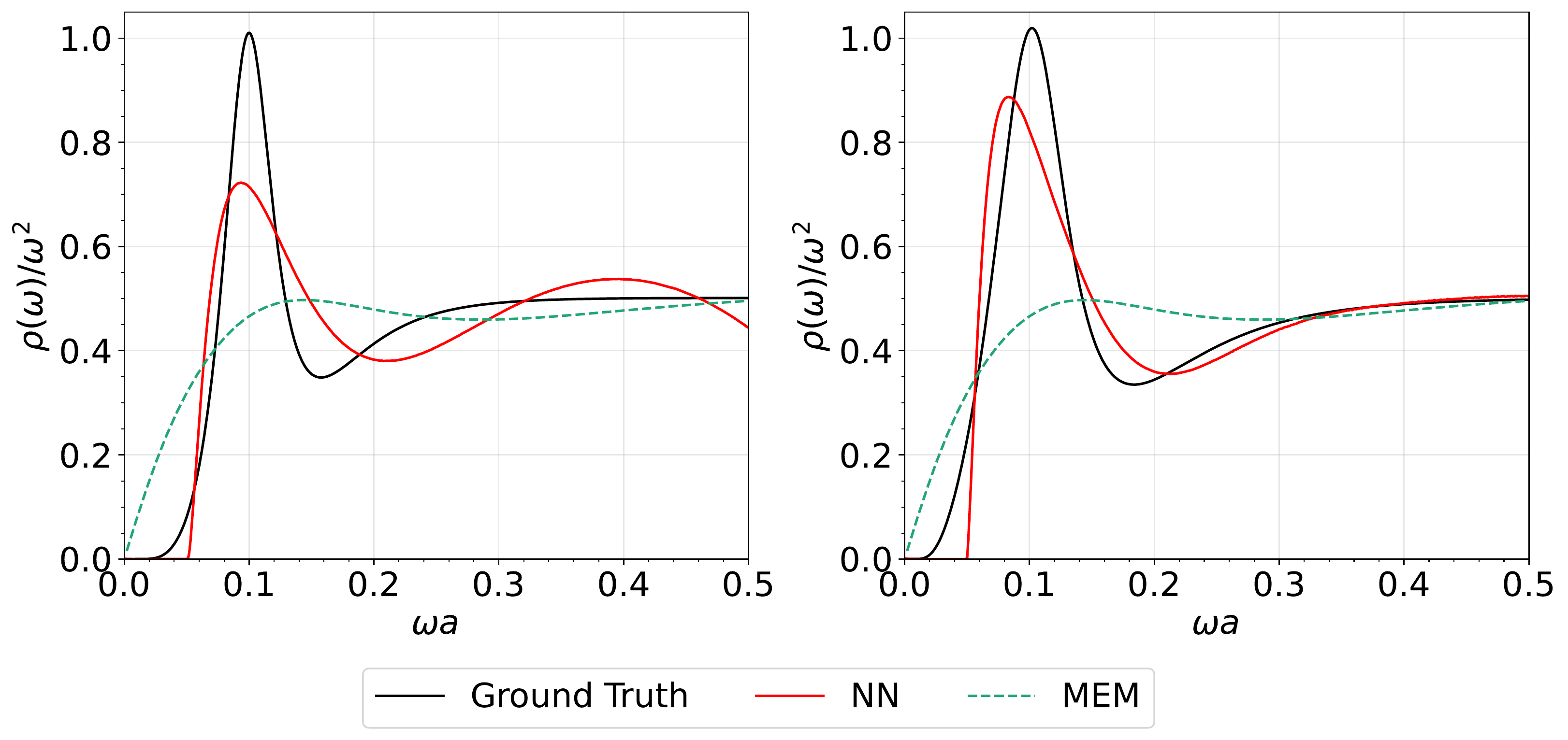}
    \caption{The predicted spectral functions from \texttt{NN} and MEM at noise level $\epsilon = 10^{-5}$ with $N_\tau = 48$.
    \label{fig:real}}
\end{figure}
We test the NN representation using two parameter sets $C_\text{res} = 2.0, C_\text{cont} = 2.1, M_\text{res} = 0.1, M_\text{cont} = 0.05$ and $\Gamma = 0.06$ (left) or  $\Gamma = 0.09$ (right). The architecture of the NN is the same as before but the multiplier factor is replaced by $\omega^2 \times\omega$ to fit the spectral behavior at the large-$\omega$ limit. In Fig.~\ref{fig:real}, MEM results lose the peak information but our reconstructions can capture it explicitly.

\emph{Summary. }
We present an automatic differentiation framework as a generic tool for unfolding spectral functions from observable data. The representations of spectral functions are with two different neural network architectures, in which non-local smoothness regularization and  modern optimization algorithm are implemented conveniently. 
We demonstrated the validity of our framework on mock examples from Breit--Wigner spectral functions with single and two peaks. To account for uncertainties from numerical simulation for the propagator observations, we confronted the framework in different levels of noise contamination for the observations. Compared to conventional MEM calculations, our framework shows superior performance especially in two peaks situation with larger noise. Also, the \texttt{NN-P2P} representation gives smooth and well-matched low frequencies spectral behavior, which is important in extracting transport properties for the system.
Owing to its ill-posedness nature, such an inverse problem cannot be fully-solved in our framework. Nevertheless, the remarkable performances of reconstructing spectral functions suggest that the framework and the freedom of introducing non-local regularization are inherent advantages of the present approach and may lead to improvements in solving the inverse problem in the future.

\textbf{Acknowledgment.---} 
We thank Drs. Heng-Tong Ding, Swagato Mukherjee and Gergely Endr\"odi for helpful discussions. The work is supported by (i) the BMBF under the ErUM-Data project (K. Z.), (ii) the AI grant of SAMSON AG, Frankfurt (K. Z. and L. W.), (iii) Xidian-FIAS International Joint Research Center (L. W), (iv) Natural Sciences and Engineering Research Council of Canada (S. S.), (v) the Bourses d'excellence pour \'etudiants \'etrangers (PBEEE) from Le Fonds de Recherche du Qu\'ebec - Nature et technologies (FRQNT) (S. S.), (vi) U.S. Department of Energy, Office of Science, Office of Nuclear Physics, grant No. DE-FG88ER40388 (S. S.). K. Z. also thanks the donation of NVIDIA GPUs from NVIDIA Corporation.

\bibliographystyle{apsrev4-1}
\bibliography{ref}

\appendix
\section{Machine learning background}
Consider that machine learning background and some of the related technical details might not be familiar to some readers, we here give a brief introduction for that. In general, most current prevalent deep learning models are implemented in automatic differentiation(AD) frameworks. AD is different from either the symbolic differentiation or the numerical differentiation~~\cite{baydin2018automatic}. Its backbone is the chain rule which can be programmed in a standard computation with the calculation of derivatives.
\begin{figure}[ht!]
    \centering
    \includegraphics[width = 0.45\textwidth]{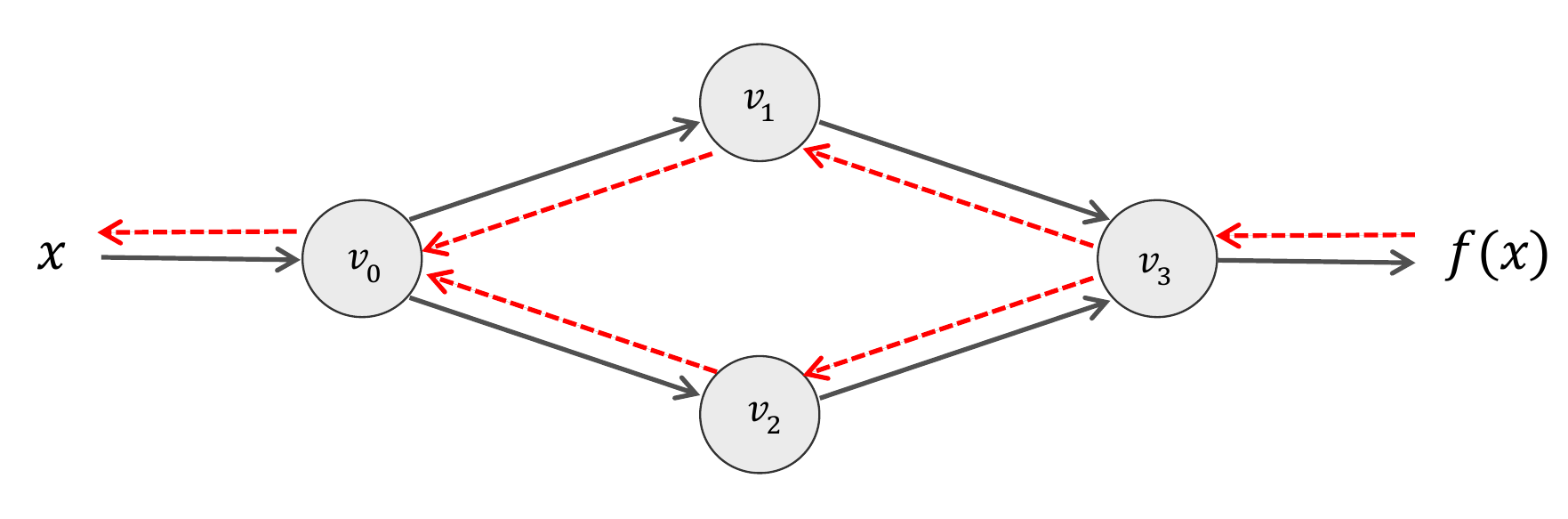}
    \caption{Computational graph of function $f(x)= \sin(x^2) + \cos(x^2)$, where $v_0(x) = x^2, v_1(v_0) = \sin(v_0), v_2(v_0) = \cos(v_0), v_3(v_1,v_2) = v_1 + v_2$. The grey solid line is the forward mode and the red dashed line indicates the reverse mode. 
    \label{fig:adsample}}
\end{figure}

 In a simplified example shown in Fig.~\ref{fig:adsample}, the computation $ y = f(x)= \sin(x^2) + \cos(x^2)$ consists of series  of differentiable operations. The forward mode is indicated by grey arrows with derivatives,
 \begin{equation}
     \dot{v}_i = \frac{\partial v_i}{\partial x}.
 \end{equation}
When calculating $f(x)$ from the input $x$, the corresponding derivatives can be evaluated by applying the chain rule simultaneously. With this intuitive example, one could understand the back-propagation(BP) algorithm~\cite{lecun2015deep} clearly. A generalized BP algorithm corresponds to the reverse mode of AD, which propagates derivatives backward from a given output. In our example, the adjoint is 
 \begin{equation}
     \bar{v}_i = \frac{\partial y}{\partial v_i},
 \end{equation}
which reflects how the output will change with respect to changes of intermediate variables $v_i$. If we treat the input $x$ as a trainable variable, the BP algorithm can be demonstrated as follows. 
\begin{align}
        \bar{v}_3&=1,\nonumber\\
        \bar{v}_1&= \bar{v}_3 \frac{\partial v_3}{\partial v_1} , \bar{v}_2= \bar{v}_3 \frac{\partial v_3}{\partial v_2} ,\nonumber\\
        \bar{v}_0&= \bar{v}_1 \frac{\partial v_1}{\partial v_0} + \bar{v}_2 \frac{\partial v_2}{\partial v_0},\nonumber\\
        \bar{x} &= \bar{v}_0\dot{v}_0.
\end{align}
The derivatives are calculated layer by layer and the final results are $\bar{x} = (\cos(x^2) - \sin(x^2))\cdot2x$. Given a target $\hat{y}$, one can define a proper loss function $\mathcal{L}(y,\hat{y})$ and fine-tune the variable $x$ with the gradient $\partial_x \mathcal{L}$, which is the well-know gradient-based optimization. The BP is crucial for training a deep neural network because the derivatives can be used to optimize a high dimensional parameter set also layer by layer~\cite{lecun2015deep}. The reverse mode of AD holds dominant advantages in the gradient-based optimization compared with the forward mode or the numerical differentiation~\cite{baydin2018automatic}.

Deep neural networks could be over-simplified as a type of compound function which has multilayer nesting structures: $f_\theta(x) = z^l(\cdots z^2(z^1(x))), z^i(z^{i-1}) = \sigma(\omega^i z^{i-1} + b^i)$, where $l$ labels the number of layers and $i$ is corresponding index, and $z^0 \equiv x $. $\sigma(\cdot)$ is a non-linear activation function and $\{\omega,b\}$ are weights and bias respectively. Weights and bias are all trainable parameters, thus could be abbreviated as $\{\theta\}$. The similar strategy explained before can be used to optimize $\{\theta\}$ with a gradient-based optimizer. As a practical example, the Adam optimizer~\cite{kingma2014adam} implemented in our work can be expressed as,
\begin{align}
    \theta^{t+1} &=  \theta^{t} - \eta \frac{\hat{m}}{\sqrt{\hat{v}} + \xi},\\
    \hat{m} &\equiv m^{t+1} = \frac{\beta_1}{1-\beta_1} m^{t} + \partial_\theta\mathcal{L}^{t},\\
    \hat{v} &\equiv v^{t+1} = \frac{\beta_2}{1-\beta_2} v^{t} + (\partial_\theta\mathcal{L}^{t})^2,
\end{align}
where the $\eta$ is learning rate, $\xi$ is a small enough scalar for preventing divergence($10^{-8}$ in our work) and $\beta_1, \beta_2$ are the forgetting factors($0.9, 0.99$ in our work) for momentum term $\hat{m}$ and its weight $\hat{v}$. The time-step $t$ labels the training step with loss function $\mathcal{L}^t(f_\theta(x))$. To ensure the neural network representation keeps some specific characters, e.g., the smoothness, one can introduce related regularization $L_s$ into the loss function to guide the training direction.

\section{Training set-ups}
In our main text, all reconstructions are learned from $N_p = 25$ points of generator in an interval $[\epsilon, 19.2 + \epsilon]$ with spacing $dp = 0.8$, where $\epsilon=10^{-3}$ is set to prevent divergence of the numerical KL kernel. For the output of neural networks, there are $N_\omega=500$ points used to represent spectral functions in an interval $0, 19.96$ with spacing $d\omega=0.04$. The same number of points are also used in the NN-P2P case. 

\begin{figure}[htpb!]
    \centering
    \includegraphics[width = 0.4\textwidth]{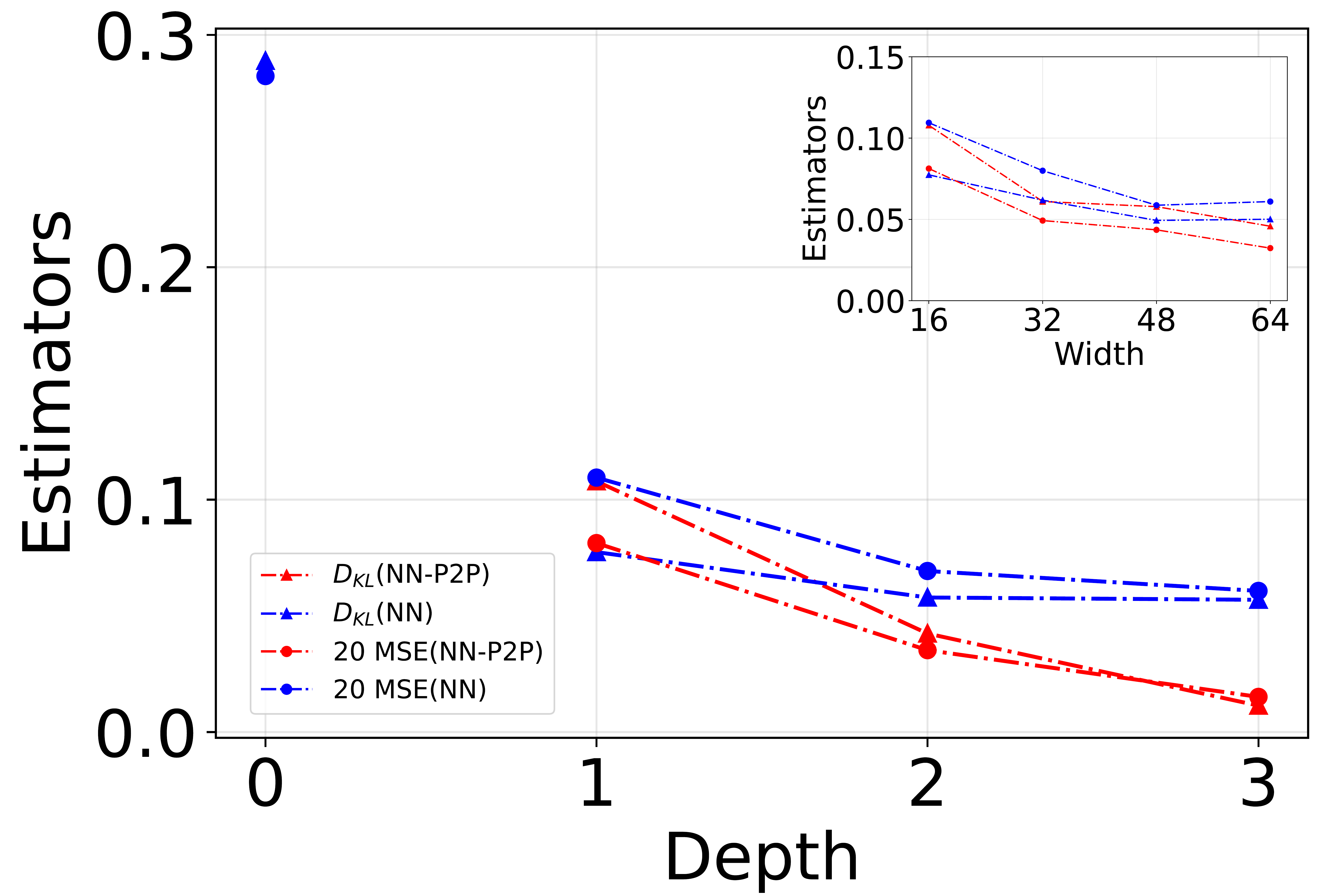}
    \caption{ Mean square error (circle) and relative entropy (triangle) with different hidden layers (width = 16), in which ``depth~=~0''  means list representation without hidden layers. The insert is estimators with different widths of neural networks (depth = 1).
    \label{fig:para}}
\end{figure}
In Fig.~\ref{fig:para}, two reconstruction performance are demonstrated for an ideal noise-free case with two network representations. The results are on the single peak case described in the manuscript. The integral form of MSE derives as,
\begin{equation}
    \int_{0}^{\infty} (\tilde{\rho}(\omega) -\rho(\omega))^2 d \omega,
\end{equation}
and the relative entropy (or Kullback--Leibler divergence) is,
\begin{equation}
    \int_{0}^{\infty} \rho(\omega) \log \left(\frac{\rho(\omega)}{\tilde{\rho}(\omega)}\right) d \omega, 
\end{equation}
where $\tilde{\rho}(\omega)$ is ground truth spectral function and $\rho(\omega)$ is the reconstructed from neural networks. The performance of two representations remain good enough after setting $\text{width} = 64$ and $\text{depth} = 3$.

\begin{table}[h!]
\caption{Training efficiency for the \texttt{NN} representations. They are set with width = 16, the same warm-up procedure and 1000 training steps.}
\centering
\def\arraystretch{1.2}
\setlength\tabcolsep{10pt}
\begin{tabular}{@{}cccc@{}}
\hline\hline
Depth & Chi-square & \multicolumn{2}{c}{Estimators}\\
\hline
  &   &MSE & DKL \\
\hline%
0 & $9.08\times 10^{-4}$ & 0.960  & 21.98\\
\hline%
1 & $4.84\times 10^{-6}$ & 0.113  & 3.572\\
\hline%
2 & $7.28\times 10^{-8}$ & 0.018  & 0.471\\
\hline%
3 & $4.89\times 10^{-8}$ & 0.028  & 1.052\\
\hline\hline

\end{tabular}
\label{tab:train}
\end{table}
In our training process, all reconstruction tasks are implemented on the machine with an Apple M1 chip through PyTorch. Each 10000 epochs cost 14s for the set-up of \texttt{NN} with width = 16 and depth = 3. When setting a same training procedure, the performances obtained by different depths of the neural network are listed in Table~\ref{tab:train}. The deeper neural network representations are more easily trained to reach a better reconstruction error than the naive list representation.

\section{Mock Lattice QCD Correlators}
In this section, we introduce the physics motivated spectral functions and their corresponding correlators in detail. The hadron spectral function $\rho(\omega,T)$ is encoded in the Euclidean correlator $G(\tau,T)$ as follows,
\begin{align}
    G(\tau,T) &= \int^\infty_0 \frac{d \omega}{2\pi}K(\omega,\tau,T)\rho(\omega,T),\label{eq:correlator}\\
    K(\omega,\tau,T) &= \frac{\cosh{\omega(\tau-\frac{1}{2T})}}{\sinh{\frac{\omega}{2T}}},\label{eq:therm}
\end{align}
where Eq.~\ref{eq:correlator} stands for a thermal correlator at temperature $T$ with the integral kernel expressed in Eq.~\ref{eq:therm}. The correlator can be calculated from Lattice QCD computations at a fixed temperature along the imaginary time $\tau$-axis. Due to the tremendous computing costs on the Lattice, there are routinely a finite number of points $N_\tau$ in $G(\tau,T)$ are available. Meanwhile, it is also limited as $T= 1/(a N_\tau)$ with the lattice spacing, $a$ which is also an energy scale in the following contents of this section. 

The physics motivated spectral function proposed in Ref.~\cite{Chen:2021giw} is used to test our framework and the results are shown in the manuscript. It can generate lattice noise-level data from a spectral function with two parts, a resonance peak and a continuum function,
\begin{align}
    \rho_\text{res}&= C_\text{res} \frac{\omega^2 }{\frac{(M_\text{res}^2 - \omega^2)^2}{M_\text{res}^2 \Gamma^2} + 1},\label{eq:res}\\
    \rho_\text{cont}&= C_\text{cont}\frac{3\omega^2}{8\pi}\theta(\omega^2 - 4M_\text{cont}^2)\tanh{(\frac{\omega}{4T})}\times\nonumber\\
    &\sqrt{1- (\frac{2M_\text{cont}}{\omega})^2}(2+(\frac{2M_\text{cont}}{\omega})^2). \label{eq:cont}
\end{align}
They are combined as,
\begin{align}
    \rho_\text{phys}(\omega,T)=&\zeta(\omega,M_\text{res},\Gamma)\rho_\text{res}(\omega,M_\text{res},\Gamma)\nonumber\\ 
    &\times(1- \zeta(\omega,M_\text{res}+\Gamma,\Gamma))\nonumber\\
    &+\zeta(\omega,M_\text{res}+\Gamma,\Gamma)\rho_\text{res}(\omega,M_\text{res},\Gamma), \label{eq:cont}
\end{align}
where $\zeta\left(\omega, M_{r e s}, \Delta\right)=1 /\left(1+e^{\frac{M_{r e s}^{2}-\omega^{2}}{\omega \Delta}}\right)$ is designed for smoothing the combination. In addition to the comparisons shown in the main text, we validate the NN reconstructions at different noise levels and(or) different numbers of correlators in Fig.~\ref{fig:real}. One different set-up should be mentioned that we prepare more correlators at the low-temperature region than the higher, which is designed to constrain the null models found in Ref.~\cite{Shi:2022yqw}. In details, for all three cases, we first prepare a same set of $N= 96$ correlators uniformly in the $\tau\in[0,1/a]$ interval. Then, for constructing $N_\tau= 16, 32, 48$ data sets, we reserve the first 11, 16 and 32 $G(\tau_i)$ and pad the rest 5, 16 and 16 uniformly from the prepared $N= 96$ correlators.
\begin{figure}[htbp!]
    \centering
    \includegraphics[width =0.48\textwidth]{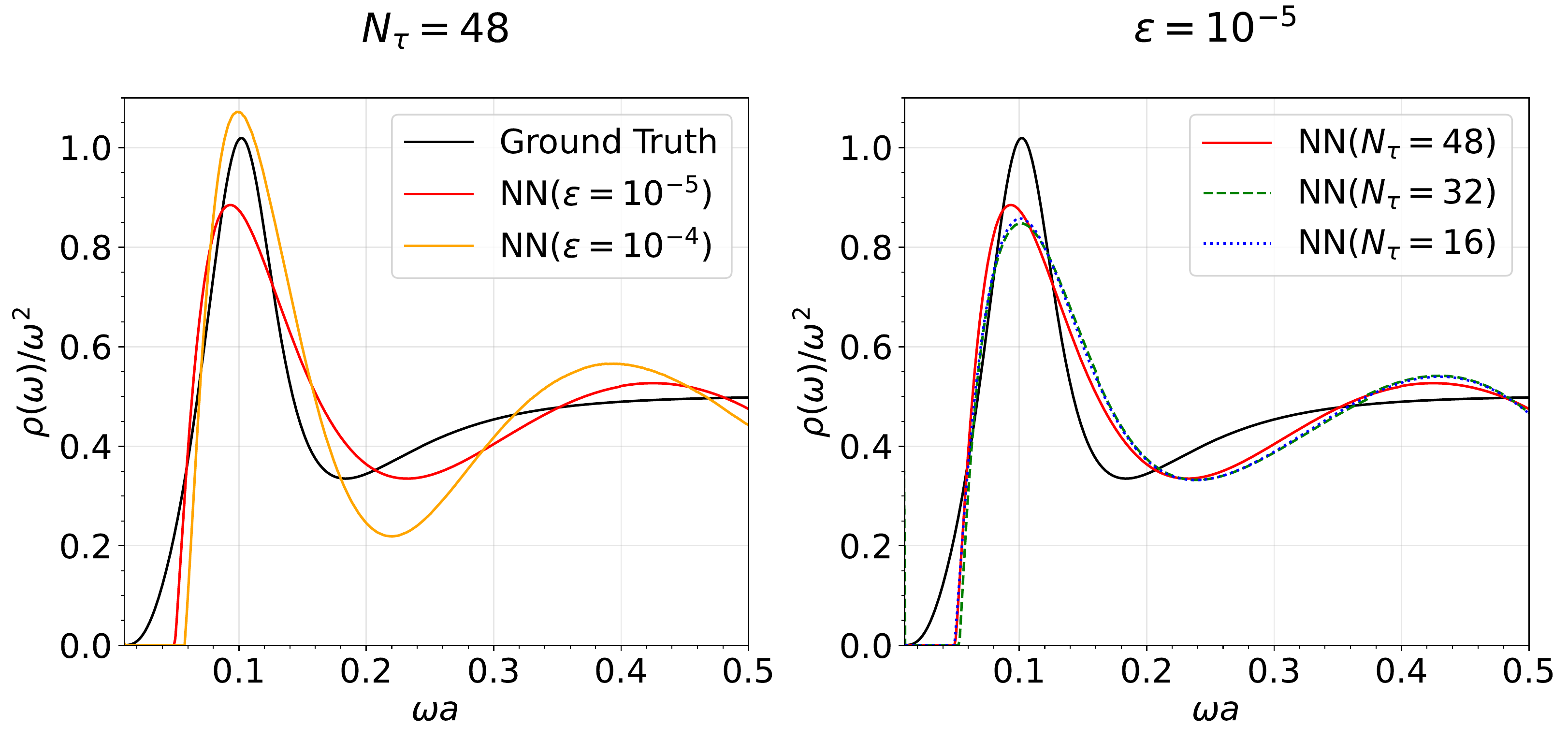}
    \caption{The predicted spectral functions from \texttt{NN} at different noise level(left panel) and different $N_\tau$(right panel).
    \label{fig:real}}
\end{figure}

\section{Detailed Comparisons with the Classical MEM}

\begin{figure}[htbp!]
    \centering
    \includegraphics[width=0.9\columnwidth]{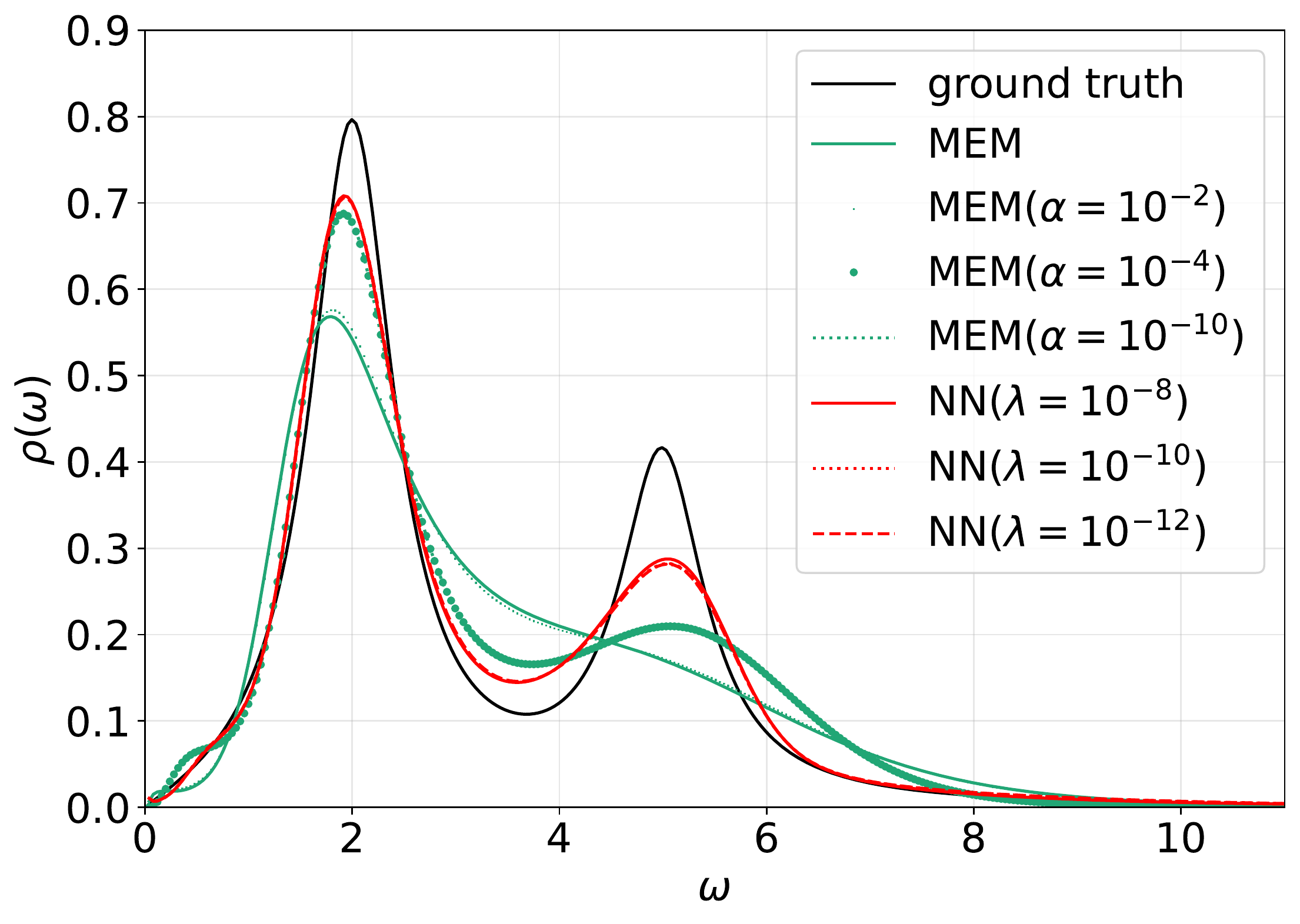}
    \caption{Comparison of \texttt{NN} and MEM with different prior coefficients for spectral function reconstruction performance at noise level $\epsilon = 10^{-4}$.}
    \label{fig: comp}
\end{figure}

As we explained in the manuscript, though the $L_2$ regularization is used to train the \texttt{NN} model, we removed the dependence and arbitrariness on the value of $\lambda$ via annealing and loosen it to small enough even zero value in the end, which is not influencing the reconstruction results of our \texttt{NN} methods, as shown in Fig.~\ref{fig: comp}. Note that as proven in our another work~\cite{Shi:2022yqw}, the uniqueness of the reconstruction holds for non-zero values of the regulator coefficient and in our manuscript we chose to use a small enough value (any value smaller than $10^{-8}$ gives the same results) for $\lambda$ in the end. So in this sense, the comparisons shown in the paper are on equal footing with respect to coefficient arbitrariness removing.
On the other hand, we tried to fix the MEM regulator coefficient with different values as well, as shown in Fig.~\ref{fig: comp}, when $\alpha$ is large the second peak of the spectral function can no be resolved, and when it's becoming smaller the best reconstruction case is also shown in Fig.~\ref{fig: comp} (as light green dotted line) but is not better than the NN model's performance.

\end{document}